\RequirePackage[hyphens]{url}
\documentclass[aip,jcp,twocolumn,showpacs,bibnotes,reprint,superscriptaddress,floatfix]{revtex4-1}

\usepackage{hyperref,url}
\usepackage{amsmath}
\usepackage[capitalise]{cleveref}
\hypersetup{hyperindex,breaklinks,colorlinks=true,linkcolor=blue,citecolor=blue,urlcolor=blue,filecolor=blue}
\usepackage{graphicx}
\usepackage{dcolumn}
\newcommand{\bk}{\mathbf{k}}

\newcommand{\bi}{\mathbf{i}}
\newcommand{\bj}{\mathbf{j}}
\newcommand{\hh}{\hat{H}}
\newcommand{\hr}{\hat{\rho}}
\newcommand{\hc}{\hat{c}}
\newcommand{\HF}{\textrm{HF}}

\begin{document}

\title{Interaction Picture Density Matrix Quantum Monte Carlo}
\author{Fionn D. Malone}
\email[Author to whom correspondence should be addressed. Electronic mail:]{f.malone13@imperial.ac.uk}
\affiliation{Department of Physics, Imperial College London, Exhibition Road, London SW7 2AZ, United Kingdom.}
\author{N. S. Blunt}
\affiliation{University Chemical Laboratory, Cambridge University, Lensfield Road, Cambridge CB2 1EW, United Kingdom.}
\author{James J. Shepherd}
\affiliation{Department of Physics, Imperial College London, Exhibition Road, London SW7 2AZ, United Kingdom.}
\affiliation{Department of Chemistry, Massachusetts Institute of Technology, Cambridge, MA, 02139, United States of America.}
\author{D.K.K. Lee}
\affiliation{Department of Physics, Imperial College London, Exhibition Road, London SW7 2AZ, United Kingdom.}
\author{J.S. Spencer}
\affiliation{Department of Physics, Imperial College London, Exhibition Road, London SW7 2AZ, United Kingdom.}
\affiliation{Department of Materials, Imperial College London, Exhibition Road, London SW7 2AZ, United Kingdom.}
\author{W.M.C. Foulkes}
\affiliation{Department of Physics, Imperial College London, Exhibition Road, London SW7 2AZ, United Kingdom.}

\begin{abstract}
    The recently developed density matrix quantum Monte Carlo (DMQMC) algorithm stochastically samples the $N$-body thermal density matrix and hence provides access to exact properties of many-particle quantum systems at arbitrary temperatures.
We demonstrate that moving to the interaction picture provides substantial benefits when applying DMQMC to interacting fermions. In this first study, we focus on a system of much recent interest: the uniform electron gas in the warm dense regime. The basis set incompleteness error at finite temperature is investigated and extrapolated via a simple Monte Carlo sampling procedure. Finally, we provide benchmark calculations for a four-electron system, comparing our results to previous work where possible.
\end{abstract}

\date{\today}
\maketitle

\section{Introduction}

The overwhelming majority of electronic structure studies of matter have been conducted at zero temperature. This state of affairs has been justified as typically one is interested in the low-energy properties of condensed matter systems or the room-temperature properties of chemical systems. Due to recent experimental advances, however, there has been renewed interest in the thermodynamic properties of electron plasmas such as those found on the pathways to inertial confinement fusion \citep{0741-3335-47-12B-S31}, in the interiors of Jupiter and other gas giants \citep{Fortney2009}, at the surfaces of solids after laser irradiation \citep{Ernstorfer20022009}, and in plasmonic catalysis \citep{doi:10.1021/nl303940z}.

Of fundamental importance to the theoretical understanding of these systems is the uniform electron gas (UEG), which has been pivotal to the development of modern quantum mechanical approaches to the low-temperature chemistry and physics of molecules and solids \citep{PhysRevLett.45.566,PhysRevB.23.5048}.
At finite temperatures the UEG can be described in terms of the density parameter, $r_s$, and the degeneracy temperature, $\Theta = T/T_F$, where $T_F$ is the Fermi temperature. When both $r_s$ and $\Theta$ are of order one the system is said to be in the warm dense regime, with quantum, thermal and interaction effects all being important. It is in this region where analytical methods, such as Ref. \citenum{PhysRevA.30.2619}, are least useful and computational approaches such as quantum Monte Carlo are most important.

In a pioneering study, Brown \emph{et al.}\ \citep{PhysRevLett.110.146405} provided the first accurate data for the UEG in the warm dense regime using the restricted path integral Monte Carlo (RPIMC) method \citep{RevModPhys.67.279}, from which the first accurate parameterizations of finite-temperature density functionals have been produced \citep{PhysRevB.88.081102,PhysRevLett.112.076403}. Recent configuration path integral Monte Carlo (CPIMC) results \citep{schoof2015ab} have called into question the validity of the restricted path approximation used in Ref. \citenum{PhysRevLett.110.146405}, with significant disagreement between the two methods at high densities and low temperatures. Simulations using a third method such as DMQMC would help to resolve this discrepancy \citep{dpimc}.

The exact thermodynamic properties of the UEG can be determined from the (unnormalized) $N$-particle density matrix
\begin{equation}
    \hat{\rho} = e^{-\beta \hh}\label{eq:dmat},
\end{equation}
where $\beta=1/k_BT$. A direct evaluation of \cref{eq:dmat} requires all eigenvalues and eigenstates of $\hh$ to be known and is an impossible task in all but the simplest of systems. The infinite basis set limit can be approached by only including determinants that can be constructed using a finite basis set of $M$ plane waves, reducing the problem to the diagonalization of an ${M\choose N }\times{M\choose N}$ matrix. Even this problem is only tractable for very small $M$ and $N$.

Recently, Booth \emph{et al.}\ have shown, through the development of the full configuration interaction QMC (FCIQMC) method, that full configuration interaction (FCI) quality results can be obtained at zero temperature with \emph{no} prior knowledge of the nodal structure of the wavefunction \citep{Booth2009}. FCIQMC also often offers a substantially reduced memory cost compared to conventional FCI calculations. This has most dramatically been seen in the case of the UEG, where a space of $\mathcal{O}(10^{108})$ Slater determinants was successfully sampled using the initiator adaptation of FCIQMC \citep{:/content/aip/journal/jcp/132/4/10.1063/1.3302277,Shepherd2012b}. Subsequently, three of us used these ideas to develop the finite-temperature analogue of FCIQMC: density matrix quantum Monte Carlo (DMQMC). This was applied to the Heisenberg model as a proof of concept \citep{PhysRevB.89.245124}, but has not previously been used to study more realistic systems.

Here we show how DMQMC can be applied to fermionic systems, starting with the UEG, thus opening the door to providing accurate, unbiased thermodynamic results for problems of chemical interest. We note that CPIMC \citep{CTPP:CTPP201100012} and Krylov-projected FCIQMC \citep{blunt2014krylov} will likely be complementary approaches to both DMQMC and PIMC in the treatment of real systems.

In \cref{sec:dmqmc} we outline the DMQMC method and show how moving to the interaction picture provides substantial improvements in statistical accuracy when treating weakly-correlated systems. In \cref{sec:basis_sets} we discuss basis-set extrapolation at finite temperatures in detail. In \cref{sec:results} we present benchmark results for a four-electron system across the relevant parameter space, comparing, where possible, to previous results. Finally, in \cref{sec:conclusions}, we outline the limitations and future prospects of DMQMC.

\section{Density Matrix Quantum Monte Carlo \label{sec:dmqmc}}

In this section we briefly outline the DMQMC algorithm; a more complete description is available in Ref.~\citenum{PhysRevB.89.245124}. DMQMC is applicable to any Hamiltonian but here we focus on the specific example of the UEG. We then explain how to sample the density matrix in the interaction picture, show that this overcomes sampling issues found when treating weakly-correlated systems, and introduce a simple Monte Carlo scheme for sampling non-interacting density matrices in the canonical ensemble. Hartree atomic units are used throughout.
\subsection{Theory}

The unnormalized density matrix in \cref{eq:dmat} obeys the symmetrized Bloch equation
\begin{align}
    \frac{d \hr}{d\beta} = -\frac{1}{2} (\hh\hr + \hr \hh ),\label{eq:bloch}
\end{align}
which can be solved using a simple Euler update scheme:
\begin{align}
    \hr(\beta+\Delta\beta) = \hr(\beta) - \frac{\Delta\beta}{2} (\hh\hr(\beta)+\hr(\beta) \hh) + \mathcal{O}(\Delta\beta^2).\label{eq:euler}
\end{align}
In DMQMC, \cref{eq:euler} is solved stochastically by evolving a population of signed `psi-particles', or `psips', in a discrete operator space made of tensor products of Slater determinants. To this end, we rewrite \cref{eq:euler} in matrix form:
\begin{align}
    \rho_{\bi\bj}(\beta+\Delta\beta) &= \rho_{\bi\bj}(\beta) - \frac{\Delta\beta}{2}\sum_\bk
        \begin{aligned}[t]
                \left[(H_{\bi\bk}-S\delta_{\bi\bk})\rho_{\bk\bj} - \right. \\
                \left. \rho_{\bi\bk}(H_{\bk\bj}-S\delta_{\bk\bj}) \right]
        \end{aligned} \\
    &= \rho_{\bi\bj}(\beta) + \frac{\Delta\beta}{2}\sum_\bk (T_{\bi\bk}\rho_{\bk\bj} + \rho_{\bi\bk}T_{\bk\bj}),\label{eq:update}
\end{align}
where $\rho_{\bi\bj} = \langle D_\bi| \hat{\rho} | D_\bj\rangle$, $|D_\bi\rangle$ is a Slater determinant in the finite but large basis set, $S$ is a variable shift introduced to control the psip population \citep{:/content/aip/journal/jcp/99/4/10.1063/1.465195,Booth2009}, and the last line defines the update matrix $T_{\bi\bj} = - (H_{\bi\bj} - S\delta_{\bi\bj})$.

The rules for evolving the psips, which resemble those used in FCIQMC \citep{Booth2009}, follow from \cref{eq:update}:
\begin{enumerate}
    \item Psips can spawn from a density matrix element $\rho_{\bi\bk}$ to $\rho_{\bi\bj}$ with probability $p_s(\bi\bk\rightarrow \bi\bj)=\Delta\beta|T_{\bk\bj}|/2$, with $\mathrm{sign}(\rho_{\bi\bj}) = \mathrm{sign}(\rho_{\bi\bk})\times\mathrm{sign}(T_{\bk\bj})$; a similar spawning process takes place from $\rho_{\bk\bj}$ to $\rho_{\bi\bj}$.
    \item Psips on the density matrix element $\rho_{\bi\bj}$ clone/die, whereby their population is increased or decreased, with probability $p_d(\bi\bj) = \frac{\Delta\beta}{2}|T_{\bi\bi}+T_{\bj\bj}|$. The population is increased if $\mathrm{sign}(T_{\bi\bi}+T_{\bj\bj})\times\mathrm{sign}(\rho_{\bi\bj}) > 0$ and decreased otherwise.
\end{enumerate}
Additionally, we annihilate psips of opposite sign on the same density matrix element to improve the efficiency of the algorithm and overcome the Fermion sign problem \citep{Booth2009,:/content/aip/journal/jcp/136/5/10.1063/1.3681396}. We note that, unlike PIMC methods where the quality of averages depends on the average sign of the sampled paths \citep{PhysRevLett.110.146405,CTPP:CTPP201100012}, in FCIQMC and DMQMC, we require a system specific and basis set dependent critical psip population to obtain correct low temperature and ground state estimates \citep{Booth2009,:/content/aip/journal/jcp/136/5/10.1063/1.3681396,:/content/aip/journal/jcp/138/2/10.1063/1.4773819,PhysRevB.90.155130}.

The simplest starting point for a simulation is at $\beta=0$, where the density matrix is the identity and can be sampled by occupying diagonal density matrix elements with uniform probability. A simulation then consists of propagating the initial distribution of psips with the rules described above to a desired value of $\beta$. Estimates for thermodynamic quantities can be found by averaging over many such simulations, a single one of which we call a `$\beta$-loop'.

\begin{figure}[h!]
    \includegraphics{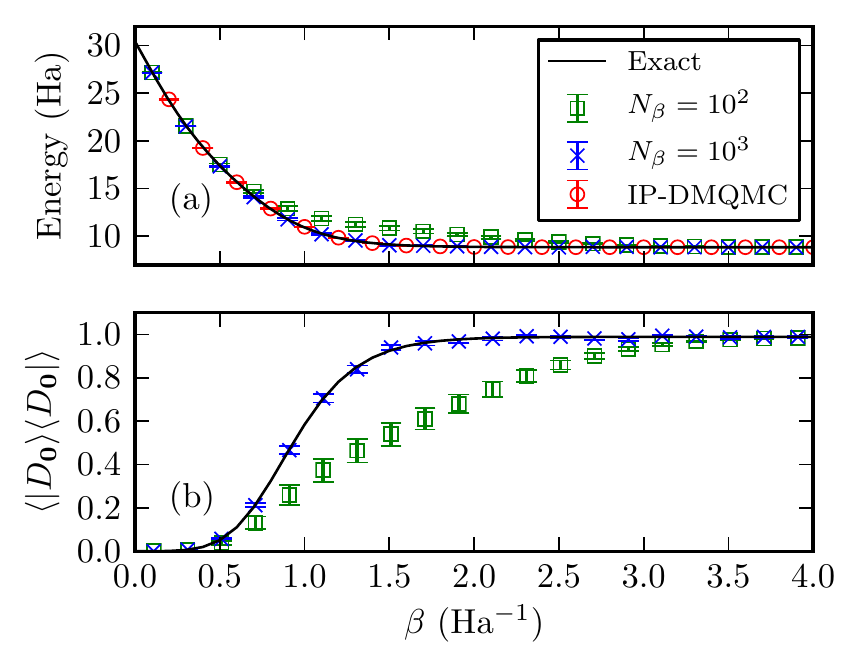}
    \caption{Panel (a) shows the total energy calculated for a seven-electron spin-polarized electron gas at $r_s=1$ with $M=33$ plane waves in the total momentum $\mathbf{K}=\mathbf{0}$ subspace using DMQMC and an initial psip population $N_p = 10^4$. Increasing the number of $\beta$-loops, $N_\beta$, from $10^2$ (squares) to $10^3$ (crosses) results in a more accurate answer being reproduced. We also see that the error bars do not reflect the true errors for $N_\beta=10^2$ in the intermediate $\beta$ regime. Panel (b) shows that the average occupation on the Hartree-Fock density matrix element ($|D_\mathbf{0}\rangle\langle D_\mathbf{0}|$) is under-represented in the intermediate temperature range. Also plotted in (a) are the IP-DMQMC results (circles) for the total energy calculated using $N_p=10^3$ and only 10 $\beta$-loops (see \cref{subsec:int_pic}).  \label{fig:dist_comp}
 }
\end{figure}

In \cref{fig:dist_comp}(a) we see that a direct application of DMQMC to the dense UEG can result in estimates for the internal or total energy that are too high in the intermediate temperature range. This can be understood by noting that, at $r_s=1$, the ground state of a few-electron UEG system is well described by a single (Hartree-Fock) determinant, $|D_\mathbf{0}\rangle$. The probability of initially selecting this determinant, however, is ${M\choose N}^{-1}$, which rapidly approaches zero as $M$ increases. If, by chance, the Hartree-Fock determinant or another low-energy determinant is sampled at $\beta = 0$, the population of psips arising from that low-energy determinant will dominate the simulation, but most simulations miss the low-energy part of the Hilbert space altogether.  As shown in \cref{fig:dist_comp}(b), this sampling problem reduces as the number of $\beta$-loops (or the population of psips per $\beta$-loop) increases, thus increasing the chance of sampling the low-energy space; however, this approach soon becomes impractical. \cref{fig:dist_comp}(a) shows that, by moving to the interaction picture, we can effectively solve this sampling issue and regain FCI-quality thermodynamic averages.

\subsection{Moving to the Interaction Picture \label{subsec:int_pic}}

There are two sampling issues present when treating real systems; the distribution of weight in the density matrix changes rapidly as a function of $\beta$, and important determinants are rarely present in our initial configurations.  Feynman originally pointed out that if we can write $\hh = \hh^0 + \hat{V}$, where $\hat{V}$ is small compared to $\hh^0$, then the quantity $e^{\beta \hh^0}\hr$ will be a slowly-varying function of $\beta$ \citep{richard1972feynman}. This does not solve the issue of selecting important determinants, so we define an auxiliary matrix
\begin{align}
    \hat{f}(\tau) = e^{-(\beta-\tau)\hat{H}^0}e^{-\tau \hh}\label{eq:fmat},
\end{align}
which has the properties
\begin{align}
    \hat{f}(\tau=0) &= e^{-\beta \hat{H}^0},\label{eq:init_cond} \\
    \hat{f}(\tau=\beta) &= e^{-\beta \hh} = \hat{\rho}(\beta)\label{eq:fin_cond}.
\end{align}
From \cref{eq:init_cond} above we see that, by working with the operator $\hat{f}$, we can start the simulation from $e^{-\beta \hat{H}^0}$ instead of the identity. For most weakly-correlated systems this should provide a good first approximation to the distribution of weight in the fully interacting density matrix.

Differentiating \cref{eq:fmat} with respect to $\tau$ we find
\begin{align}
    \frac{d \hat{f}}{d\tau} &= \hat{H}^0\hat{f} - \hat{f}\hh \label{eq:int_prop}\\
                                          &= -\hat{V}_I(\tau-\beta)\hat{f}\label{eq:int_pic},
\end{align}
where we have used the usual definition of an operator in the interaction picture:
\begin{align}
    \hat{A}_I(\tau) = e^{\tau \hat{H}^0}\hat{A}e^{-\tau\hat{H}^0}.
\end{align}
In practice the exponential factors appearing in \cref{eq:int_pic} are time-consuming to evaluate and we prefer to work with \cref{eq:int_prop}. Since we choose $\hat{H}^0$ to be diagonal in our determinantal basis set of Hartree-Fock eigenstates, the only modification to the original DMQMC algorithm is that $p_d(\bi\bj) = \Delta \tau |H^0_{\bi\bi} - H_{\bj\bj}|$. This scheme, which we dub interaction-picture DMQMC (IP-DMQMC), has the added benefit that there is typically $\emph{no}$ death along the diagonal as long as \mbox{$H^0_{\bi\bi}\ge H_{\bi\bi}$}. This overcomes a related issue in DMQMC simulations of large systems, whereby the weight of walkers on the diagonal decays nearly to zero with $\beta$; this was previously remedied with the use of importance sampling \cite{PhysRevB.89.245124}.

\cref{eq:fin_cond} shows that IP-DMQMC only samples the correct distribution at $\tau =\beta$ so that, unlike in DMQMC, where the whole temperature range is sampled in one simulation, separate simulations are required for each $\beta$ value. As in DMQMC, estimates for observables require averaging over multiple independent simulations.

Referring back to \cref{fig:dist_comp} we see that working in the interaction picture effectively eliminates this sampling issue, with the correct total energy being reproduced using small numbers of psips and $\beta$-loops.

\subsection{Sampling the initial condition\label{sec:init_cond}}
The choice of $\hat{H}^0$ is somewhat arbitrary, but it should allow for an efficient sampling of $\hat{f}(\tau=0)$ and this is most easily achieved if $\hat{H}^0$ is non-interacting. In principle, any initial density matrix can be sampled using the Metropolis algorithm \citep{:/content/aip/journal/jcp/21/6/10.1063/1.1699114}, but we have found this approach problematic due to the long equilibration times required at low temperatures and in large basis sets. An alternative method, which is free from such issues, is to use knowledge of the grand canonical density matrix corresponding to $\hat{H}^{0}$ and sample this in such a way that the desired, canonical, distribution is reached.

Consider the grand canonical density matrix
\begin{align}
    \hr^0_{GC} &= e^{-\beta(\hh_0-\mu \hat{N})},
\end{align}
where
$\hh_0 = \sum_i \varepsilon_i \hat{c}^{\dagger}_i \hat{c}^{}_i$ is some non-interacting Hamiltonian whose single-particle eigenvalues $\varepsilon_i$ are known, $\hat{N}$ is the number operator and $\mu$ is the chemical potential, which can be determined from the implicit relationship
\begin{align}
    \langle \hat{N} \rangle_{GC} &= \sum_i \frac{1}{e^{\beta(\varepsilon_i-\mu)}+1}\label{eq:cpot}\\
                           &= \sum_i p_i,
\end{align}
where we have identified the usual Fermi-factor, $p_i$.
The grand canonical partition function, $Z_{GC}$, can be written as
\begin{align}
    Z_{GC} &= \sum_N\sum_{\{n_i^N\}} e^{-\beta\sum_i(\varepsilon_i -\mu)n_i}
\end{align}
where $\{n_i^N\}$ denotes a set of occupation numbers such that $\sum_{i} n_i = N$ and $n_i \in \{0,1\}$ for fermions.

\begin{figure}
    \includegraphics{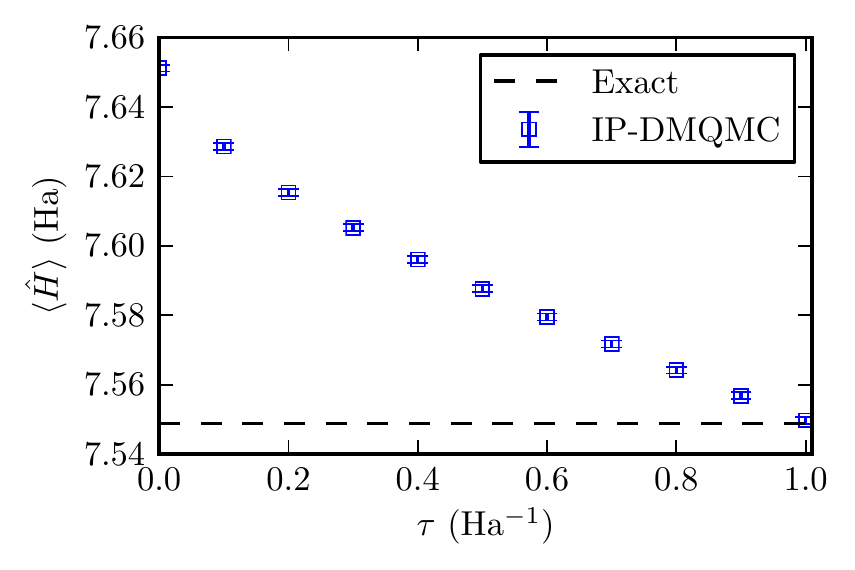}
    \caption{Variation of $\langle \hh\rangle$ with $\tau$ using $\hat{H}^0=\sum_\bk \varepsilon_\bk \hc^{\dagger}_\bk\hc^{}_\bk$, i.e., the free-electron expression with $\varepsilon_\bk = \frac{1}{2}\bk^2$. The grand canonical procedure described in \cref{sec:init_cond} was used. The system shown is a four-electron, spin polarized gas at $r_s =1$, $M=81$ and $\beta = 1$. For these results we used approximately $10^3$ psips and averaged over 100 simulations. The dashed line represents the exact FCI result, which IP-DMQMC reproduces at $\tau=\beta$ as expected. For comparison $\langle \hh \rangle = 50.751(4)$ Ha at $\beta = 0$.}
    \label{fig:slow_var}
\end{figure}

The probability of selecting a particular set $\{\bar{n}^N_i\}$ is
\begin{align}
    P_{GC}(\{\bar{n}^{N}_i\}) = \frac{1}{Z_{GC}}\prod_{n_i \in \{\bar{n}^N_i\}} e^{-\beta(\varepsilon_i-\mu) n_i}.
\end{align}
However, we wish to generate determinants in the canonical ensemble where the correct probability is
\begin{align}
P_C(\{\bar{n}^N_i\}) = \frac{1}{Z_C}\prod_{n_i \in \{\bar{n}^N_i \}} e^{-\beta(\varepsilon_i-\mu)n_i}\label{eq:can_prob},
\end{align}
and
\begin{align}
    Z_C = \sum_{\{n_i^N\}} e^{-\beta\sum_i(\varepsilon_i-\mu)n_i}.
\end{align}
We see that $P_{C}(\{n_i^N\}) \propto P_{GC}(\{n_i^N\})$. Thus, by independently occupying orbitals with probability $p_i$ and then discarding those configurations with $\langle \hat{N} \rangle \ne N$, we attain the correct proportionality factor $Z_{GC}/Z_C$. Only about one in $\sqrt{N}$ of the configurations sampled has the right value of $N$, but the sampling process is so fast that very little computer time is required for any system size in the reach of many-body simulation methods. The chemical potential can be obtained by numerically inverting \cref{eq:cpot} in the appropriate finite basis set, a procedure we carry out using Scipy \citep{jonescipy}.
A demonstration of the above procedure is given in \cref{fig:slow_var}, where we see that $\langle \hh\rangle$ is indeed a slowly varying function of $\tau$ and that the correct estimate is reproduced at $\tau=\beta$.

Finally, we note that any diagonal density matrix can be obtained by reweighting the configurations which result from the above sampling procedure as
\begin{equation}
    \label{eq:reweighting}
    P_{\mathrm{new}}(\{\bar{n}_i^N\}) = P_{\mathrm{old}}(\{\bar{n}_i^N\}) e^{-\beta (E_{\mathrm{new}} - E_\mathrm{old})},
\end{equation}
where $E_{\mathrm{new}}$ and $E_{\mathrm{old}}$ are the new and old total energies of a given configuration $\{\bar{n}_i^N\}$, respectively.

\section{Basis set extrapolation\label{sec:basis_sets}}

To treat the UEG using DMQMC we need to work in a finite basis set of $M$ plane waves; thermodynamic quantities will therefore need to be extrapolated to the $M \rightarrow \infty$ limit. At $T=0$ it has been found \citep{Shepherd2012b,Shepherd2012a,PhysRevB.86.035111} that the correlation energy, $E_c = E - E_{\HF}$, for unpolarized systems converges like $M^{-1}$. Recent CPIMC results \citep{CTPP:CTPP201400072} were also obtained using an $M^{-1}$ extrapolation, although in principle this relationship only holds for an unpolarized system; we have found that, for polarized systems, extrapolating with $M^{-5/3}$ results in a better fit \citep{mp2sppp}, consistent with similar observations by other authors \citep{PhysRevB.90.075125,:/content/aip/journal/jcp/141/16/10.1063/1.4900447}.  Based on the analysis presented here, Schoof \emph{et al.} \citep{schoof2015ab} have used the $M^{-5/3}$ extrapolation with CPIMC. In addition, we find that the convergence of the total energy is strongly dependent on temperature.

\begin{figure}
    \includegraphics{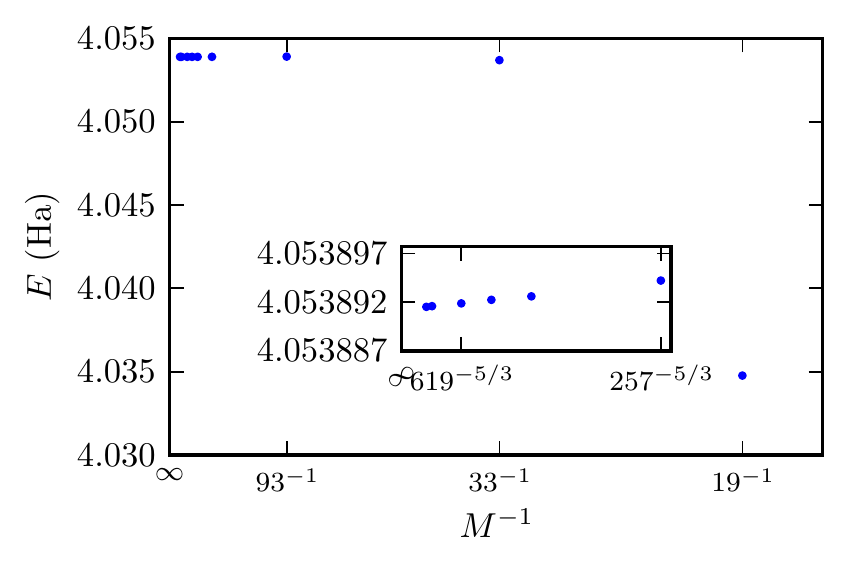}
    \caption{Behavior of the FCI total energy with basis-set size for an $N = 2$, $r_s = 1$ spin-polarized system at $\Theta=0.5$. Here we see the competition between the exponential convergence of the total energy at low $M$ (main plot) and the $M^{-5/3}$ behavior for high $M$ (inset).}
    \label{fig:2e_exc}
\end{figure}

At $T>0$ there is a competition between the convergence of the kinetic and potential energies with $M$. To investigate this further we focus on a two-electron spin-polarized system, which can be solved exactly using diagonalization in large basis sets. In \cref{fig:2e_exc} we see this competition between energy scales: the total energy initially increases rapidly with basis-set size before appearing to saturate. As the size of the basis set is further increased, a slight reduction in the total energy is observed, with the residual error apparently proportional to $M^{-5/3}$ (see inset of \cref{fig:2e_exc}). In this regime the convergence of the total energy is dominated by exchange and correlation effects.

The initial increase of the total energy with respect to $M$ at non-zero temperatures can be understood by looking at the non-interacting total energy as a function of basis-set size, which is most easily analyzed in the grand canonical ensemble.  The non-interacting basis-set error is
\begin{align}
    \Delta E_0(M) &= E_0(\infty) - E_0(M)\\
             &= \sum_{\bk >\bk_c} \varepsilon_\bk p_\bk,
\end{align}
where $k_c$ is the plane-wave cutoff. For $\varepsilon_c = \frac{1}{2} k_c^2 \gg 1$, this can be approximated as
\begin{align}
    \Delta E_0 \approx \int_{\varepsilon_c}^{\infty} \varepsilon^{3/2} e^{-\beta(\varepsilon - \mu)} d\varepsilon
\end{align}
where we have used $p_{\bk} \approx e^{-\beta(\varepsilon_\bk - \mu)}$ for $\varepsilon_\bk \gg \mu$.
Hence
\begin{align}
    \Delta E_0(M) &\approx \int_0^{\infty} (\varepsilon_c+x)^{3/2} e^{-\beta(\varepsilon_c+x-\mu)} dx\\
                  &\approx \varepsilon_c^{3/2} e^{-\beta(\varepsilon_c -\mu)}\int_0^{\infty} e^{-\beta x} dx,
\end{align}
if $\varepsilon_c \gg \beta^{-1}$ so that $(\varepsilon_c+x)^{3/2} \approx \varepsilon_c^{3/2}$ everywhere $e^{-\beta x}$ is significant.
It then follows that the leading-order correction is
\begin{align}\label{eq:asymp}
    \Delta E_0(M) &\approx \beta^{-1} \varepsilon_c^{3/2} e^{-\beta(\varepsilon_c-\mu)}.
\end{align}
From \cref{eq:asymp} we see that the kinetic energy begins to converge exponentially once $\varepsilon_c \approx \beta^{-1}$ or $M \approx V(\frac{\Theta}{r_s^2})^{3/2} \approx N \Theta^{3/2}$, where $V$ is the simulation-cell volume. In practice, we find that for large $\Theta$ the kinetic energy and hence the total energy converge quite slowly. This is an issue for DMQMC simulations as the cost of a calculation increases with basis-set size.

\begin{figure}
    \includegraphics{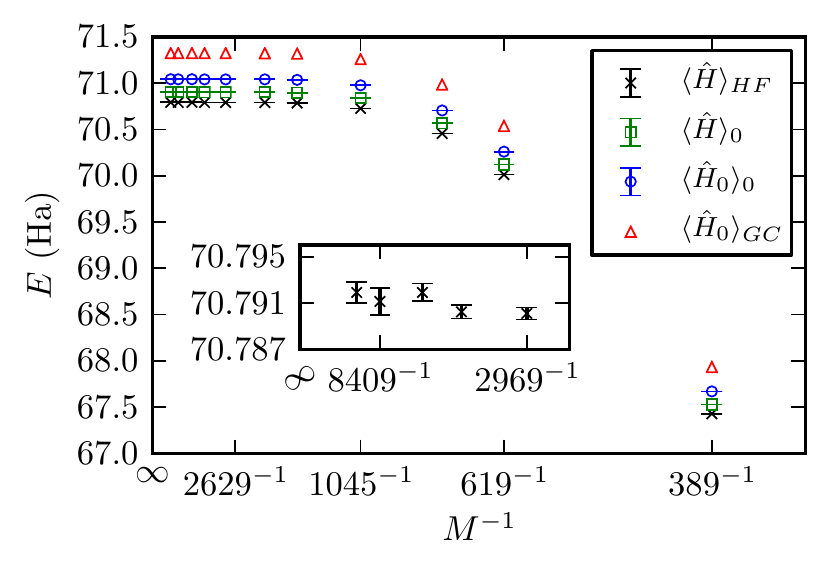}
    \caption{Exponential convergence of the Hartree-Fock total energy with basis-set size for $N = 4$, $r_s = 1$ and $\Theta = 4$. Note that no Madelung contribution is included for the Hartree-Fock estimates in this figure. The infinite basis-set limit for $E_{\HF}$ is estimated as 70.792(1) Ha. For comparison, we plot $\langle \hh \rangle_0$ and $\langle \hh_0\rangle_0$, where the trace is now with respect to the non-interacting density matrix. Also plotted is the total energy calculated in the grand canonical ensemble using a finite basis set, i.e., $\langle \hat{H}_0\rangle_{GC} = \sum_\bk^{\bk_c} \varepsilon_\bk p_\bk$.}
    \label{fig:kinetic}
\end{figure}

We can mitigate some of these issues by instead extrapolating the temperature-dependent correlation energy, $E_c(\beta, M) = E(\beta, M) - E_{\HF}(\beta, M)$. The infinite basis-set total energy can then be reconstructed as \mbox{$E(\beta, M=\infty) = E_{\HF}(\beta, M=\infty) + E_c(\beta, M=\infty)$}. We calculate the `Hartree-Fock' energy, \mbox{$E_{\HF} = \langle \hh \rangle_{\HF}$}, using the density matrix \citep{fthf}

\begin{equation}
    \hat{\rho}_{\HF} = \sum_\bi e^{-\beta E_\bi^{\HF}} |D_\bi \rangle \langle D_\bi |\label{eq:hf_dmat},
\end{equation}
where $E_\bi^{\HF} = \langle D_\bi |\hat{H}|D_\bi\rangle$ and the sum runs over all determinants in the basis set. $E_{\HF}(\beta)$ can be found using the sampling procedure outlined in \cref{sec:init_cond}, i.e.,
\begin{align}
    E_{\HF} &= \frac{1}{Z_{\HF}} \sum_\bi E^{\HF}_{\bi}e^{-\beta E^{\HF}_{\bi}}\\
           &= \frac{1}{Z_{\HF}} \sum_\bi E^{\HF}_\bi e^{-\beta (E^{\HF}_{\bi}-E^0_{\bi})} e^{\beta E^0_\bi}\\
           &= \frac{\sum_\bi E^{\HF}_{\bi} w(\bi) p(\bi)}{ \sum_\bi w(\bi) p(\bi)},
\end{align}
where $w(\bi) = e^{-\beta(E^{\HF}_{\bi}-E^0_\bi)}$ and $p(\bi) = Z_0^{-1} e^{-\beta E^0_\bi}$. Thus, by generating determinants as described in \cref{sec:init_cond} and reweighting them using $w(\bi)$, we can instead sample $\hat{\rho}_{\HF}$ and, as a result, estimate $E_{\HF}$ as desired. In \cref{fig:kinetic} we show the convergence of $E_{\HF}(\beta,M)$ as a function of basis set for a four-electron, spin-polarized system at $r_s=1$ and $\Theta=4$. Note the large basis-set sizes required to converge the total energy to within statistical error bars. \cref{fig:kinetic} also shows various other `non-interacting' or `mean-field' energy estimates as functions of $M$. Any of these could in principle be subtracted from $E(\beta,M)$ to define a correlation energy, but the quantity defined by subtracting $E_{\HF}(\beta,M)$  extrapolates most smoothly to the infinite $M$ limit. The non-interacting grand canonical energy, $\langle \hat{H}_0 \rangle_{GC}$, is significantly larger than the canonical estimates.

\begin{figure}
    \includegraphics{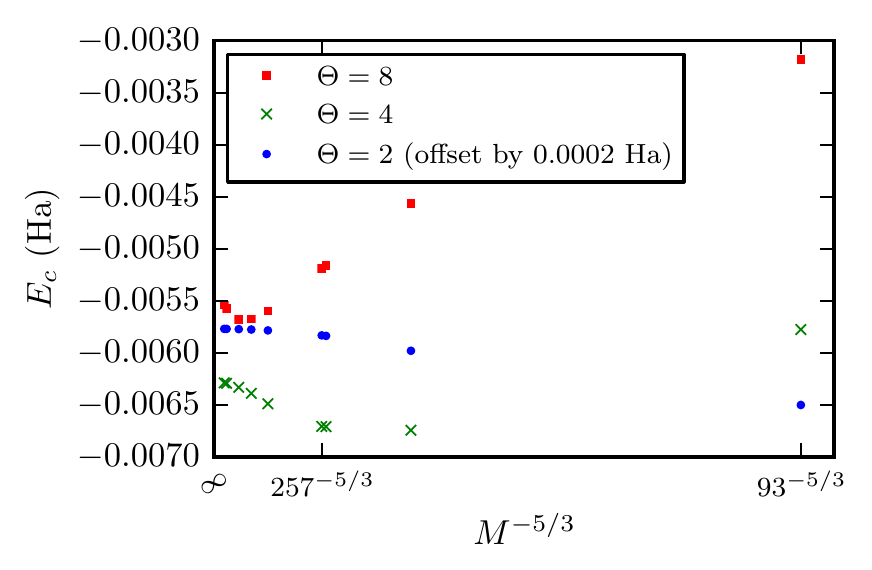}
\caption{Comparison of the convergence of $E_c$ with basis-set size calculated using exact diagonalization for a two-electron spin-polarized system at $r_s=1$ for different values of $\Theta$. The $\Theta=2$ data has been shifted down by 0.0002 Ha for visibility. The dependence of $E_c$ on $\Theta$ is non-monotonic. The converged-basis-set correlation energies at $\Theta = 8$ and $\Theta=2$ are
very similar, but the correlation energy at $\Theta = 4$ is more negative.}
    \label{fig:2e_extrap}
\end{figure}

\cref{fig:2e_extrap} shows how $E_c(\beta,M)$ depends on $M$ at a number of different temperatures. For small basis sets $E_c$ shows a power-law decay with $M$, but this ceases for large enough $M$ and the energies begin to increase again.
We believe that the increase is due to kinetic effects that are not captured in the non-interacting expression we subtract. The value of $M$ at which the correlation energy begins to increase again corresponds to the onset of the power-law convergence of the total energy observed in \cref{fig:2e_exc}. This non-variational behavior of the internal energy with respect to $M$ is not surprising as, at finite temperatures, it is the free energy that satisfies a variational principle.

We can estimate $E_c(\beta,M=\infty)$ by taking the value calculated with the largest basis set after the minimum is reached. This will in general over estimate $E_c$ (it will be too negative) but the remaining discrepancy is typically smaller than the stochastic error bar. The systematic errors left after extrapolating $E_c(\beta,M)$ in this manner are well within chemical accuracy \mbox{($\sim k_BT$)} and can be orders of magnitude smaller than those introduced by a direct extrapolation of the total energy.

\section{Results \label{sec:results}}

\subsection*{Computational Methods}

All calculations discussed in this paper were performed using the HANDE code \citep{HANDE}.
Unless otherwise stated, the QMC calculations used real amplitudes \citep{PhysRevLett.109.230201,overy2014unbiased,:/content/aip/journal/jcp/142/18/10.1063/1.4920975} to sample the density matrix, which improves stochastic efficiency compared to integer weights.
The full data set is available in the supplementary material \citep{supp}.

\subsection{Four electron uniform electron gas}
\begin{figure}[h!]
    \includegraphics{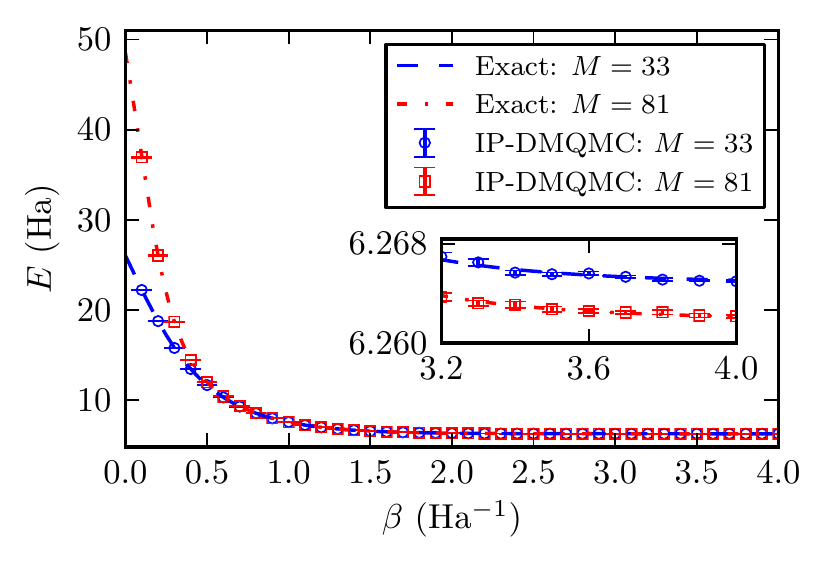}
    \caption{Comparison of IP-DMQMC results (markers) with FCI results (dashed-lines) for the UEG with $N = 4$ and $r_s=1$ in two different basis-sets. The inset shows the low $T$ behavior where we see increasing the basis-set size serves to decrease the total energy in contrast to the high $T$ behavior, where the opposite occurs. These calculations used integer rather than real weights and were run using approximately 1000 psips and we averaged over 100 simulations.}
    \label{fig:exact_comp}
\end{figure}
Using the procedures outlined above we are now in a position to provide exact benchmarks for the UEG in small simulation cells across the relevant parameter space. In this first study we focus on the four-electron spin-polarized system, which is the smallest non-trivial system and one for which there already exist benchmark calculations \citep{CTPP:CTPP201400072}. All energies contain the Madelung contribution \citep{PhysRevB.53.1814} where appropriate. As a first step we compare our four-electron IP-DMQMC results to FCI results in small basis sets and see perfect agreement across the whole temperature range (\cref{fig:exact_comp}).

We have extended these results to basis sets far beyond the reach of conventional full diagonalization procedures; the largest space sampled here contains approximately $10^{22}$ density matrix elements. We used the initialization procedure outlined in \cref{sec:init_cond} and the free-electron Hamiltonian for $\hh^0$ for $r_s \le 1$; for $r_s > 1$ we found it advantageous to use Hartree-Fock density matrix defined in \cref{eq:hf_dmat}. The calculations were initialized with $10^3$--$10^7$ psips and the results averaged over $100$--$5000$ simulations, each using a different random number seed. Time steps $\Delta\tau$ ranging from $0.01/E_F$ to $0.001/E_F$ were used, with a smaller time-step required at lower $r_s$; the values chosen were small enough that we could resolve no time-step error within the statistical errors. Each $(r_s,\Theta, M)$ calculation was typically run for 2 hours on 48 cores with a total computational cost of approximately 80000 core hours. The separate calculations of $E_{\HF}$ required 9000 core hours.

\begin{figure}[h!]
    \includegraphics{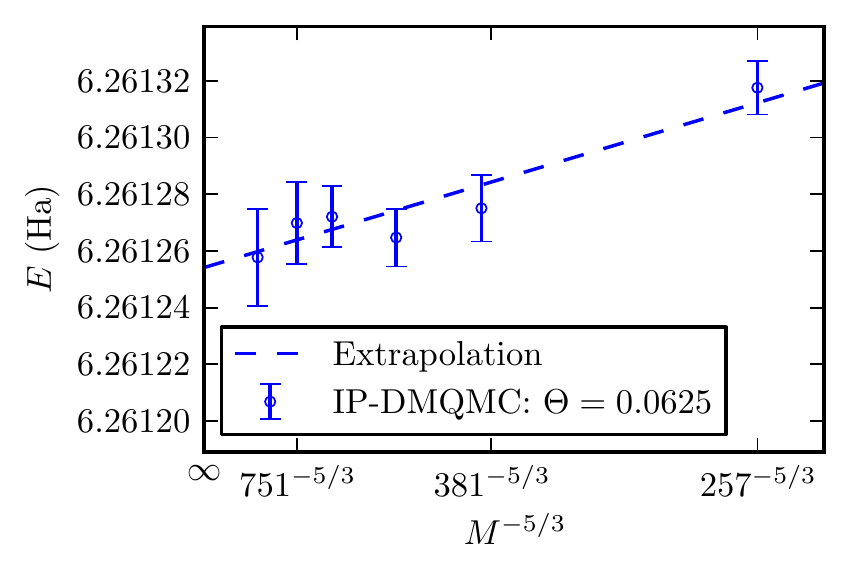}
    \caption{Total energy of a four-electron spin-polarized system at $r_s=1$ and $\Theta=0.0625$, showing a convergence with $M^{-5/3}$. The dashed line represents an extrapolation to the infinite-basis-set limit carried out using a least-squares fit as implemented in Scipy \citep{jonescipy}. At this low temperature, a direct extrapolation of the total energy works better than the correlation-energy extrapolation technique discussed in \cref{sec:basis_sets}.}
    \label{fig:4e_exc}
\end{figure}

\begin{figure}[h!]
    \includegraphics{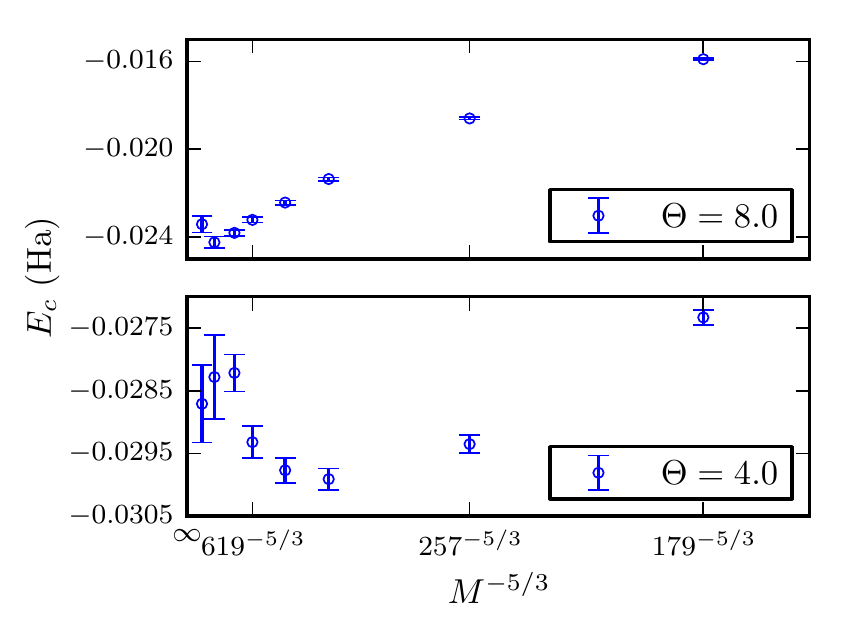}
    \caption{Convergence of $E_{c}$ with basis-set size for the four-electron system at $r_s=5$ calculated using IP-DMQMC. Note the similar behaviour to that found in the two-electron case in \cref{fig:2e_extrap}. At the high temperatures considered here, the correlation-energy extrapolation technique introduced in \cref{sec:basis_sets} works much better than the total-energy extrapolation illustrated in \cref{fig:4e_exc}.} 
    \label{fig:4e_elowT}
\end{figure}

\begin{figure}[h!]
\includegraphics{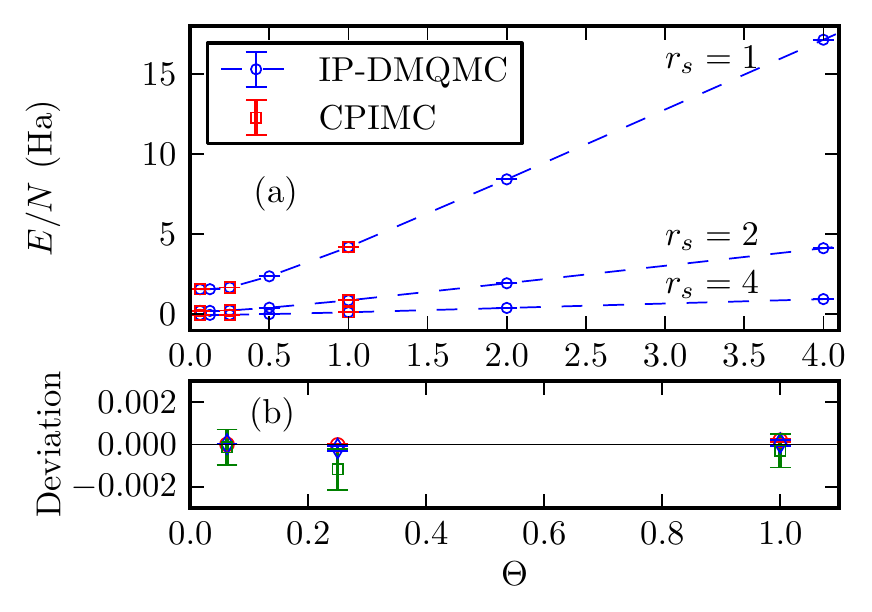}
\caption{(a) Extrapolated total energies per particle for the four-electron system at $r_s = 1, 2, 4$ showing exact agreement with the CPIMC results of Ref.~\citenum{CTPP:CTPP201400072}. Dashed lines are meant as guides to the eye. (b) Relative deviation, ($E_{\mathrm{DMQMC}}-E_{\mathrm{CPIMC}})/E_{\mathrm{DMQMC}}$, as a function of temperature showing statistically identical results for $r_s = 1 $ (circles), $r_s=2$ (diamonds) and $r_s=4$ (squares). Further results at higher temperatures and other $r_s$ values are available in tabular form in the supplementary material and again agree with the available CPIMC results.}
\label{fig:cpimc_comp}
\end{figure}

In \cref{fig:4e_elowT,fig:4e_exc} we show the convergence of the IP-DMQMC results with basis set at low and high temperatures, respectively. We note that the behavior found in the two-electron system is also found in the four-electron system; in particular the non-trivial dependence of the correlation energy on $M$ is reproduced. We find that a direct extrapolation of the total energy with respect to $M$ is best for $\Theta \le 0.25$ as it is here where kinetic effects are minimal and there is a clear trend in the total energy for the basis set sizes considered. The procedure outlined in \cref{sec:basis_sets} is best suited for temperatures above this, becoming increasingly useful for $\Theta \ge 2$ as more highly excited states become accessible, requiring prohibitively large basis sets for a direct extrapolation to be possible. In between these too regimes both methods produce statistically identical results\citep{supp}. \cref{fig:cpimc_comp} summarizes our results and shows perfect agreement with the available CPIMC data from Ref.~\citenum{CTPP:CTPP201400072}. Further results at higher temperatures and other $r_s$ values are available in tabular form in the supplementary material and again agree with the available CPIMC results.
\section{Discussion and Conclusions \label{sec:conclusions}}
In this paper we have demonstrated how DMQMC can be applied to realistic systems. By moving to the interaction picture we have removed sampling issues found when treating weakly-correlated systems with large basis sets.

We have examined in detail the convergence of the total and correlation energies with respect to basis-set size $M$ and temperature using a system accessible to FCI calculations. We found that, in general, these quantities exhibit a non-trivial dependence on $M$ attributable to the competing energy scales present. By developing a simple Monte Carlo sampling scheme, we showed that it is possible to reduce the error in extrapolating these quantities to the complete basis-set limit by at least an order of magnitude at high temperatures. We believe this analysis and developments will be useful when treating molecular systems at finite temperatures. In addition, our approach to calculating the `Hartree-Fock' energy should a useful first approximation when providing accurate benchmark calculations for systems away from the thermodynamic limit and in analyzing single-particle finite-size effects for the UEG at non-zero temperatures.

Using these developments we have reproduced the four-electron CPIMC benchmarks of Ref.~\citenum{CTPP:CTPP201400072} and provided results at higher temperatures. We hope that these small-system results will aid the analysis of the apparent discrepancies between other QMC methods for larger system sizes \citep{full_comp} and serve as benchmarks for other QMC methods based in configuration space.

Whilst the results presented here are for much smaller systems than those accessible by RPMIC and CPIMC, DMQMC provides access to exact finite-temperature data for a given basis set.  The main limitation on the system size is the critical population (determined by the plateau height\cite{PhysRevB.89.245124,:/content/aip/journal/jcp/136/5/10.1063/1.3681396}) required to sample the density matrix.  There are several grounds for optimism.  The sign problem is much weaker at higher temperatures, implying that larger systems will be accessible albeit over a restricted temperature range.
Further, in our previous study \citep{PhysRevB.89.245124} we found that the plateau height in DMQMC, which is a measure of the strength of the sign problem, was roughly the square of that in FCIQMC, which would suggest a rather limited utility of our method. Remarkably, however, we have found that, for the UEG at various $r_s$ values and for some other models, the plateau height is only a small \emph{multiple} of the FCIQMC plateau height. For example, for $N=4$, $r_s=5$ and $M=1045$, the FCIQMC plateau height is roughly 8000 psips in comparison to 90000 in IP-DMQMC at $\Theta=0.0625$. Finally, there is evidence that methodological developments will increase the scope of systems that can be treated with DMQMC. Given that the initiator approximation\cite{:/content/aip/journal/jcp/132/4/10.1063/1.3302277} enabled FCIQMC to be applied successfully to large UEG systems across a range of densities \citep{Shepherd2012a, Shepherd2012b}, we are confident that, using similar approximations and developments in importance sampling, DMQMC will become a competitive method in treating degenerate Fermi systems.

\begin{acknowledgments}
FDM is funded by an Imperial College PhD Scholarship. NSB acknowledges Trinity College, Cambridge for funding. JJS thanks the Royal Commission for the Exhibition of 1851 for a Research Fellowship. JSS and WMCF acknowledge the research environment provided by the Thomas Young Centre under Grant No.~TYC-101. WMCF received support from the UK Engineering and Physical Sciences Research Council under grant EP/K038141/1. Computing facilities were provided by the High Performance Computing Service of Imperial College London, by the Swiss National Supercomputing Centre (CSCS) under project ID s523, and by ARCHER, the UK National Supercomputing Service, under EPSRC grants EP/K038141/1 and via a RAP award.
\end{acknowledgments}

\bibliography{refs}

\begin{thebibliography}{42}
\bibitem [{dpi()}]{dpimc}%
  \BibitemOpen
  \href@noop {} {}\bibinfo {howpublished} {There has also been disagreement
  reported at high densities between RPIMC and direct path integral Monte Carlo
  \citep{PhysRevE.91.033108}, although the origin of these discrepancies
  remains unclear.}\BibitemShut {Stop}%
\bibitem [{mp2()}]{mp2sppp}%
  \BibitemOpen
  \href@noop {} {}\bibinfo {howpublished} {This power law can be derived by
  analyzing the MP2 correlation energy of a polarized system in a manner
  analogous to the treatment of the unpolarized electron gas in
  Ref.~\citenum{PhysRevB.86.035111}.}\BibitemShut {Stop}%
\bibitem [{fth()}]{fthf}%
  \BibitemOpen
  \href@noop {} {}\bibinfo {howpublished} {This differs from the usual approach
  to finite-temperature Hartree-Fock theory \citep{Mermin196399}, in which the
  density matrix is obtained by minimizing the free energy over a choice of all
  non-interacting trial density matrices, with the single-particle eigenvalues
  and (in general) orbitals allowed to depend on temperature. Here, in place of
  a sum of single-particle eigenvalues, we use the many-particle total energies
  of specific determinants in our Boltzmann factors. Fortunately, since our
  `Hartree-Fock' density matrix is used only to improve the efficiency of the
  IP-DMQMC algorithm, the simulations yield exact results whether or not the
  standard definition is used.}\BibitemShut {Stop}%
\bibitem [{ful()}]{full_comp}%
  \BibitemOpen
  \href@noop {} {}\bibinfo {howpublished} {A full comparison will be presented
  elsewhere.}\BibitemShut {Stop}%
\end{thebibliography}

\end{document}